\documentclass[fleqn,usenatbib]{mnras}

\usepackage{txfonts}

\usepackage[T1]{fontenc}
\usepackage{ae,aecompl}

\usepackage{psfig}   
\usepackage{graphicx}
\usepackage{amssymb}
\usepackage{subfigure}

\title[Planet--planet binaries in star-forming regions]{Can planet--planet binaries survive in star-forming regions?}

\author[]{
  Richard J.~Parker\thanks{E-mail: r.parker@sheffield.ac.uk}\thanks{Royal Society Dorothy Hodgkin fellow}, Simon P. Goodwin and Jessica L. Diamond \vspace*{0.1cm}\\
   Astrophysics Research Cluster, School of Mathematical and Physical Sciences, The University of Sheffield, Hounsfield Road, Sheffield, S3 7RH}

\begin{document}

\date{}
                             
\pagerange{\pageref{firstpage}--\pageref{lastpage}} \pubyear{2025}

\maketitle

\label{firstpage}

\begin{abstract}
Significant numbers of free-floating planetary-mass objects have been discovered in nearby star-forming regions by the James Webb Space Telescope, including a substantial number (42) of Jupiter Mass Binary Objects (`JuMBOs') in the Orion Nebula Cluster. The JuMBOs have much wider separations than other populations of substellar binaries, and their existence challenges conventional theories of substellar and planetary-mass object formation. Whilst several theories have been proposed to explain their formation, there has yet to be a study that determines whether they could survive the dynamical encounters prevalent within a dense star-forming region. We place a population of planet--planet binaries in $N$-body simulations of dense star-forming regions and calculate their binary fraction over time. We find that between 50 -- 90\,per cent of planet--planet binaries are destroyed on timescales of a few Myr, which implies that many more must form if we are to observe them in their current numbers. Furthermore, if the ONC was much more dense at formation, the initial separation distribution of the JuMBOs must have been even wider (and less similar to other substellar binaries) than the observed distribution.   
\end{abstract}


\section{Introduction}

Understanding the boundary between stars and planets (or whether a boundary exists) is one of the most important topics in modern astrophysics, and is being revolutionised by the first data from the James Webb Space Telescope (JWST). In particular, JWST has discovered significant numbers of free-floating planetary-mass objects \citep{McCaughrean23,Langeveld24,Luhman24a,Luhman24b,Luhman24c}.

Some of these substellar objects appear to be in binaries \citep[Jupiter-Mass Binary Objects, or `JuMBOs' for short,][]{Pearson23} with projected separations 28 -- 384\,au. These systems are unusually wide for substellar binaries \citep[which typically have separations below 10\,au,][]{Burgasser07,Thies07,Factor23}, and pose a challenge for formation theories.

The fidelity of the JuMBOs has been questioned by \citet{Luhman24a} who finds that their colours are more consistent with reddened background sources, rather than young substellar objects in a star-forming region. In spite of this uncertainty, several theories have been proposed to explain them \citep{Lazzoni24,Zwart24}, ranging from liberation of pairs of planets from their parent stars \citep{Wang24} to photoerosion of the cores of systems that would otherwise go on to form much wider stellar-mass binaries \citep{Whitworth04,Diamond24}.

The candidate JuMBOs were discovered in the Orion Nebula Cluster (ONC), the most dense star-forming region within 500\,pc of the Sun \citep{King12a}, and many studies have demonstrated that stellar and substellar binaries are disrupted in star-forming regions with densities $\geq$100\,M$_\odot$\,pc$^{-3}$ \citep{Kroupa99,Parker11a,Parker11c,Marks12,Parker23d}.

In this Letter, we investigate whether planet--planet binary systems can survive in dense star-forming regions, and what the implications of this are for the JuMBOs observed with JWST. We present our methods in Section~\ref{methods}, we present our results in Section~\ref{results} and we conclude in Section~\ref{conclusions}. 

\section{Methods}
\label{methods}

We run $N$-body simulations of the dynamical evolution of star-forming regions in which we place a population of planet--planet binaries. The regions contain $N_{\rm sys} =  1500$ systems -- either a single star or a planet-planet binary system. For simplicity, the stellar systems are all single stars\footnote{In the ONC the stellar binary fraction is similar to that in the Galactic field, $\sim$50\,per cent \citep{King12a}. If we were to include stellar binaries, the number of destructive encounters would likely be higher, due to the higher collisional cross section of the binaries compared to single stars \citep{Li15,Li20}.}, with masses drawn from a \citet{Maschberger13} Initial Mass Function, which has a probability distribution for selecting a mass $m$ of the form
\begin{equation}
p(m) \propto \left(\frac{m}{\mu}\right)^{-\alpha}\left(1 + \left(\frac{m}{\mu}\right)^{1 - \alpha}\right)^{-\beta}.
\label{maschberger_imf}
\end{equation}
In this equation, $\alpha = 2.3$ is the \citet{Salpeter55} slope describing the high-mass end of the IMF, and $\beta = 1.4$. $\mu = 0.2$\,M$_\odot$, and we adopt an upper limit to the IMF of $m_{\rm up} = 50$\,M$_\odot$. In most of our simulations (A, B, D, E) we adopt a lower mass limit of $m_{\rm low} = 0.08$\,M$_\odot$, i.e. we do not create a population of brown dwarfs that overlap in mass with the the JuMBOs. However, in one set (C) we adopt $m_{\rm low} = 0.01$\,M$_\odot$ to create a population of single brown dwarfs.

We randomly select 10\,per cent of the systems to be planet--planet binaries, and randomly draw their component masses from the JuMBO masses provided in \citet{Pearson23}. In three sets of simulations (A, B, C), we draw the separations of these binaries randomly from the JuMBO catalogue (we take the separation from the same system from which we draw the component masses), but in two sets (D and E) we draw them from a flat distribution between 50 - 500\,au. 

We do not have any information on the eccentricity of the observed JuMBO systems, and we therefore set all eccentricities to zero.

\begin{table}
\caption[bf]{Summary of simulation set-ups. The columns show the simulation label, the lower limit to the initial mass function for the single systems, $m_{\rm low}$, the separation distribution for the planet--planet binaries, and the initial median local stellar density $\tilde{\rho}$ in the star-forming region.}
\begin{center}
\begin{tabular}{|c|c|c|c|}
\hline 
Sim. & $m_{\rm low}$ & planet--planet separations  & $\tilde{\rho}$  \\
\hline
A & 0.08\,M$_\odot$ & JuMBOs (28 -- 384\,au) & 10\,000\,M$_\odot$\,pc$^{-3}$ \\
B & 0.08\,M$_\odot$ & JuMBOs (28 -- 384\,au) & 100\,M$_\odot$\,pc$^{-3}$ \\
\hline
C & 0.01\,M$_\odot$ & JuMBOs (28 -- 384\,au) & 10\,000\,M$_\odot$\,pc$^{-3}$ \\
\hline
D & 0.08\,M$_\odot$ & 50 -- 500\,au & 10\,000\,M$_\odot$\,pc$^{-3}$ \\
E & 0.08\,M$_\odot$ & 50 -- 500\,au & 100\,M$_\odot$\,pc$^{-3}$ \\
\hline
\end{tabular}
\end{center}
\label{cluster_sims}
\end{table}

Observations \citep[e.g.][]{Gomez93,Cartwright04,Sanchez09,Hacar13,Andre14} and simulations \citep[e.g.][]{Schmeja06,Bate09,Girichidis11,Dale12b} suggest that stars form in filamentary structures, which then converge to form hubs of star formation \citep{Myers11}, resulting in a spatially substructured distribution for the stars.

A convenient way of setting up $N$-body simulations with substructure is to use the box fractal method \citep{Goodwin04a}, which has the advantage that the degree of spatial and kinematic substructure is described by just one number, the fractal dimension $D$. For a detailed description of the fractal set-up, we refer the interested reader to \citet{Goodwin04a} and \citet{DaffernPowell20}. In three dimensions, a fractal with a high degree of substructure is created with $D = 1.6$, whereas a uniform sphere is produced when $D = 3.0$. Observed star-forming regions are all consistent with having evolved from  much more substructured distributions \citep{DaffernPowell20} and we therefore adopt $D = 1.6$ in all of our simulations.

The velocities of the objects in the fractal distribution are set so that the velocity dispersion is small on local scales, but can be quite different on larger scales, similar to observations \citep{Larson81,Hacar13,Henshaw16a}.

Observations and simulations also suggest that young stars are likely to have a low velocity dispersion with respect to the gravitational potential, i.e.\,\,be subvirial \citep{Foster15,Kuznetsova15,Vazquez19}, and we scale the velocties to be subvirial ($\alpha_{\rm vir} = 0.3$, where the virial ratio $\alpha_{\rm vir} =  T/|\Omega|$, and $T$ and $\Omega$ are the total kinetic and potential energies of the objects, respectively). 

We select two radii, $r_F$ for the star-forming regions, such that our simulations encompass the full range of possible initial densities for the ONC. The present-day stellar density in the central regions of the ONC is several hundred M$_\odot$\,pc$^{-3}$ \citep{King12a}, and it is possible that the initial density was similar. However, `reverse engineering' -- the process of comparing properties of $N$-body simulations (such as the amount of spatial substructure, number of runaway and walkaway stars, degree of mass segregation, binary star orbital distributions)  -- suggests a much higher initial density, perhaps up to $10^4$\,M$_\odot$\,pc$^{-3}$ \citep{Allison10,King12a,Marks12,Parker14b,Parker14e,Schoettler20}. To account for these two extremes, we adopt $r_F = 1$\,pc, and $r_F = 5$\,pc, which for the fractal dimension $D = 1.6$ and $N_{\rm sys} = 1500$ results in densities of $10^4$\,M$_\odot$\,pc$^{-3}$ and 100\,M$_\odot$\,pc$^{-3}$, respectively.

We evolve the simulations using the \texttt{kira} Hermite $N$-body integrator within the \texttt{Starlab} environment \citep{Zwart99,Zwart01} for 10\,Myr, such that we comfortably exceed the current estimates of the age of the ONC \citep[1 -- 4\,Myr, even taking into account potential age spreads,][]{Jeffries11,Reggiani11b,Bell13,Beccari17}. We do not include stellar evolution in the simulations, nor do we include a background gas potential.

A summary  of the different initial conditions is shown in Table~\ref{cluster_sims}. We run ten versions of each simulation, identical apart from the random number seed used to initialise the stellar masses, positions and velocities. The results presented in the subsequent figures are the average of the ten versions of each simulation.

\section{Results}
\label{results}

\begin{figure}
  \begin{center}
\setlength{\subfigcapskip}{10pt}
\rotatebox{270}{\includegraphics[scale=0.35]{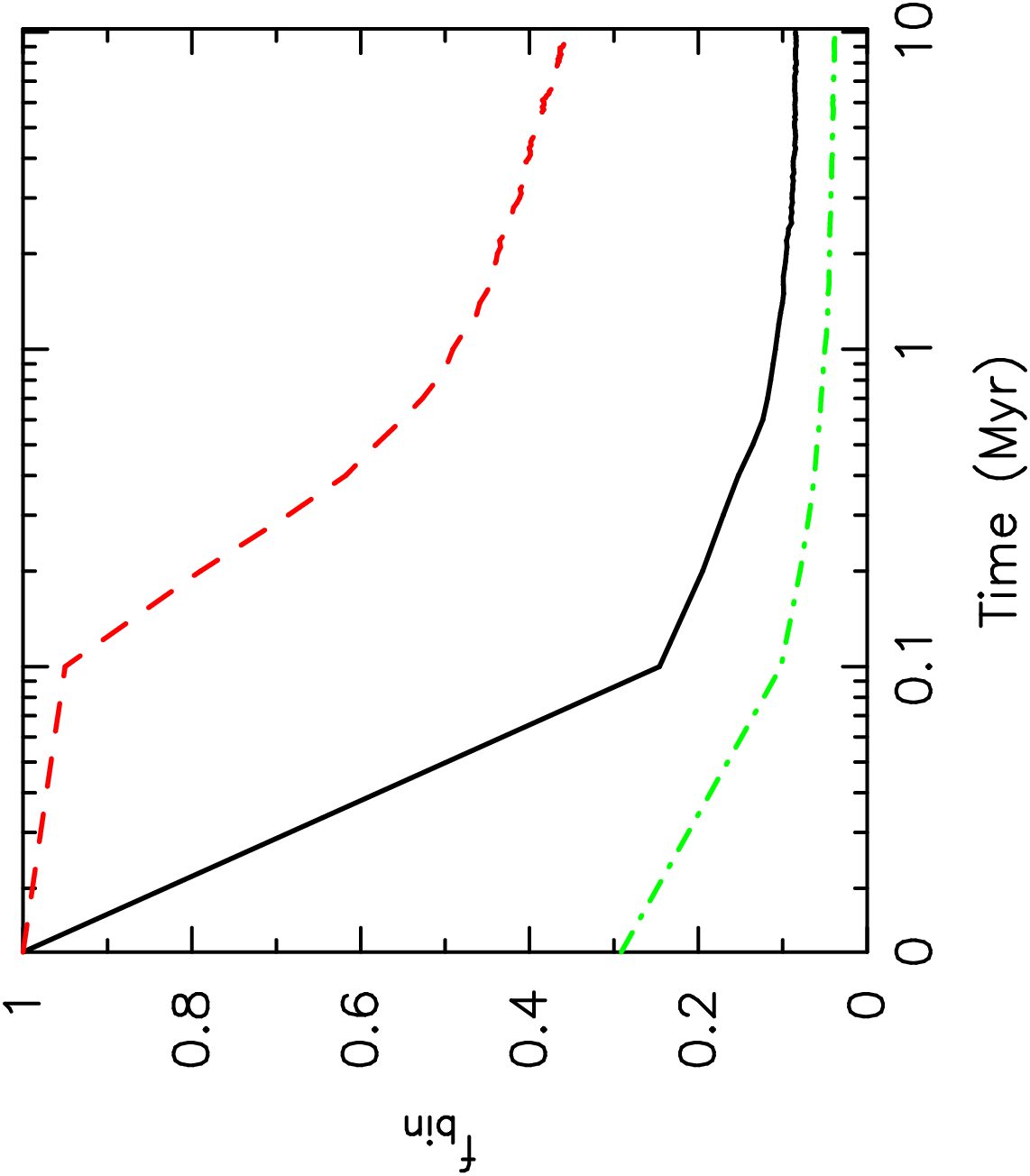}} 
\hspace*{0.3cm} 
\caption[bf]{Evolution of the substellar-mass binary fraction (as defined in Eqn.~\ref{eqn:fbin}) in three of our simulations. The solid black line shows the evolution of the binary fraction in our dense simulations (A). The red dashed line shows the evolution of the binary fraction in the lower-density simulations (B). The green dot-dashed line shows the evolution of the binary fraction in dense simulations that include a population of brown dwarfs drawn from the initial mass function, and so the initial binary fraction is $\sim 0.3$ (C). In all three of these simulations, the planet--planet separations are drawn from the JuMBO catalogue, and we have averaged together the results from ten realisations of the same initial conditions. }
\label{fig:fbin}
  \end{center}
\end{figure}

The relatively wide planet--planet binaries, and their relatively small binding energies, make them susceptible to disruption in all of our simulated star-forming regions. Fig.~\ref{fig:fbin} shows the evolution of the binary fraction of low-mass objects, which is defined as 
\begin{equation}
  f_{\rm bin} = \frac{B}{S + B},
  \label{eqn:fbin}
\end{equation}
where $S$ is the number of singles, and $B$ is the number of binary systems. The lines shown in Fig.~\ref{fig:fbin} are for systems below the hydrogen-burning mass limit (0.08\,M$_\odot$), so do not include stellar-mass objects.

The solid black line shows the binary fraction in the dense star-forming regions where we draw the planet--planet properties from the JuMBO distributions (sim. A). In 1\,Myr the destruction of planet--planet binaries is such that the binary fraction reduces from unity to 0.1. Even in the less dense star-forming regions (sim. B), the destruction of systems is significant, with the binary fraction reducing from unity to 0.5 in 1\,Myr (and continuing to decrease throughout the remainder of the simulation), as shown by the red dashed line.

The green dot-dashed line shows a simulation (C) where the planet--planet binary properties are taken from the observed JuMBO distribution, but where the simulations include other substellar mass objects down to 0.01\,M$_\odot$. This produces an initial binary fraction of 0.29, but the destruction of the planet--planet binaries reduces this fraction to 0.05 after 1\,Myr. For clarity, we only show the simulations where the separations are drawn from the observed JuMBO distribution (A, B, C). The binary fractions in simulations where the separations are drawn from a flat distribution between 50 -- 500\,au (D and E) evolve in a similar manner to simulations A and B.

It is clear from these simulations that a significant proportion of the observed JuMBOs would not survive in a star-forming region with densities commensurate with the majority of nearby star-forming regions (i.e. $\geq 100$\,M$_\odot$\,pc$^{-3}$). Or, to invert the statement, there must have been many more planet--planet binaries with similar properties, that have since been destroyed therough dynamical encounters, than the $\sim$40 JuMBOs presently observed in the ONC. 

\begin{figure}
  \begin{center}
\rotatebox{270}{\includegraphics[scale=0.35]{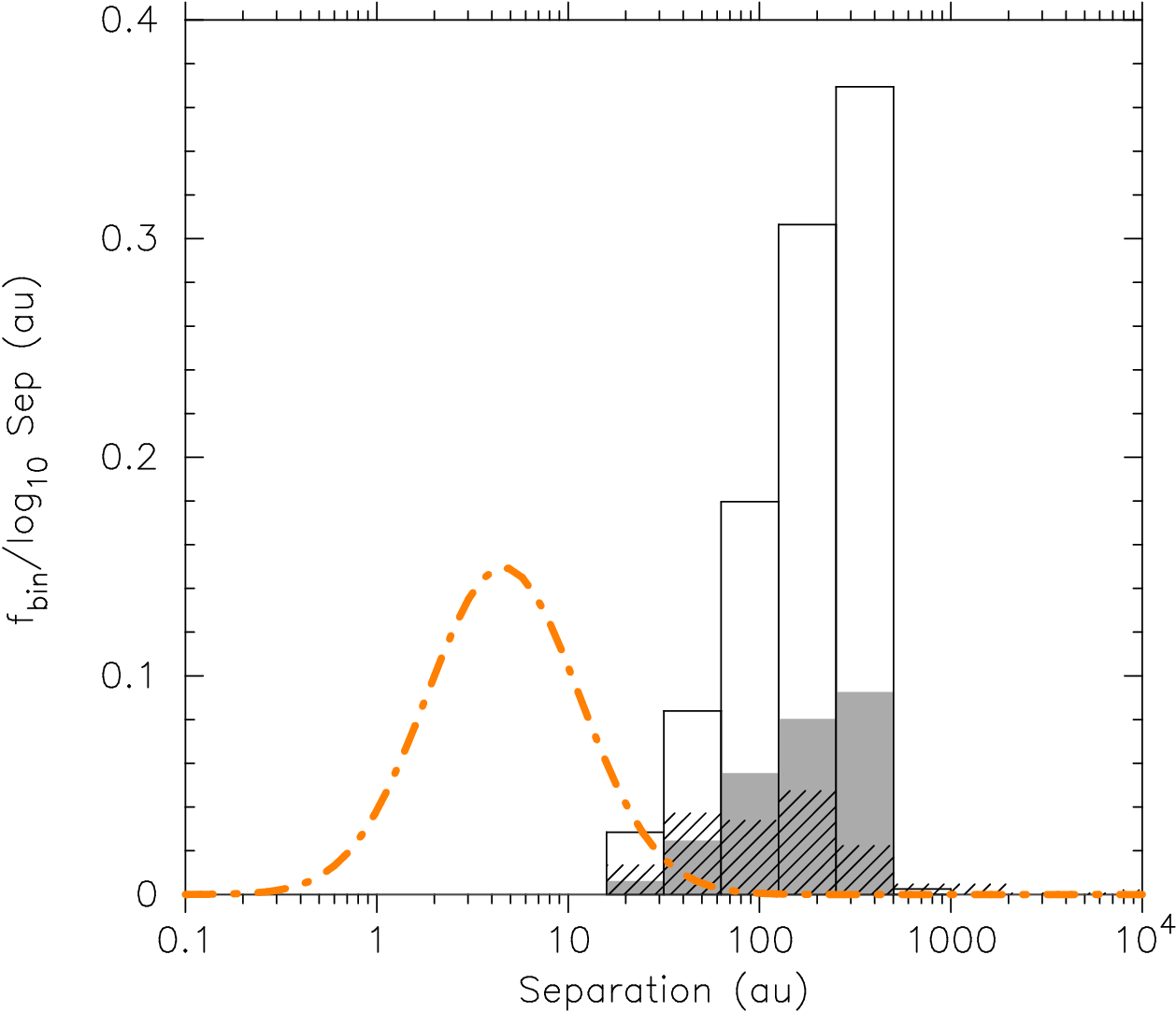}}
\caption[bf]{The evolution of the planet--planet binary separation distribution in $N$-body simulations. The open histogram shows the initial population in our $N$-body simulations, and the hashed histogram showns the population after 1\,Myr of dynamical evolution. The solid grey histogram is the observed JuMBO distribution \citep{Pearson23}, and the orange dot-dashed line is the fit to brown dwarf-brown dwarf binaries in the nearby Galaxy \citep{Basri06,Burgasser07,Thies07}.}
\label{fig:PP_sepdist}
  \end{center}
\end{figure}

We now examine the effects of this dynamical destruction on the planet--planet binary separation distribution. In Fig.~\ref{fig:PP_sepdist} we show histograms of the planet--planet binaries from our simulation  (C) with an initial substellar binary fraction of 0.29. The open histogram is the distribution at 0\,Myr, which is drawn from the observed JuMBO distribution (the solid grey histogram), but scaled upwards to reflect the many more systems we place in the simulation. The hatched histogram is the distribution after 1\,Myr of dynamical evolution in the $N$-body simulation. For reference, the fit to the substellar binary distribution observed in the nearby Galaxy is shown by the orange dot-dashed line, and peaks towards much smaller separations \citep{Basri06,Burgasser07,Thies07,Factor23}.

\begin{figure*}
  \begin{center}
\setlength{\subfigcapskip}{10pt}
\hspace*{-1.5cm}\subfigure[$\tilde{\rho} = 10^4$M$_\odot$\,pc$^{-3}$]{\label{fig:PP_sepcum-a}\rotatebox{270}{\includegraphics[scale=0.35]{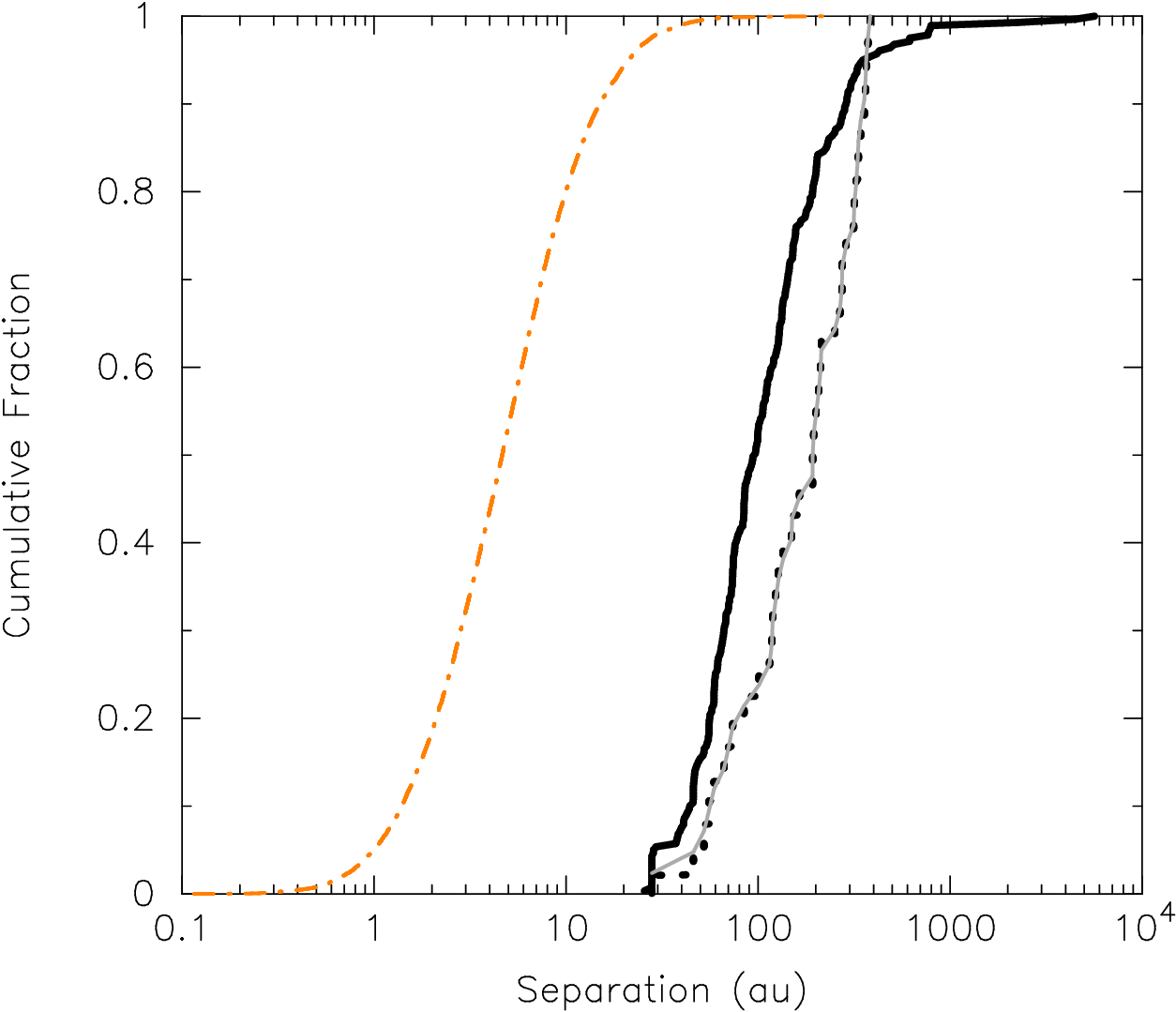}}} 
\hspace*{0.3cm} 
\subfigure[$\tilde{\rho} = 10^2$M$_\odot$\,pc$^{-3}$]{\label{fig:PP_sepcum-b}\rotatebox{270}{\includegraphics[scale=0.35]{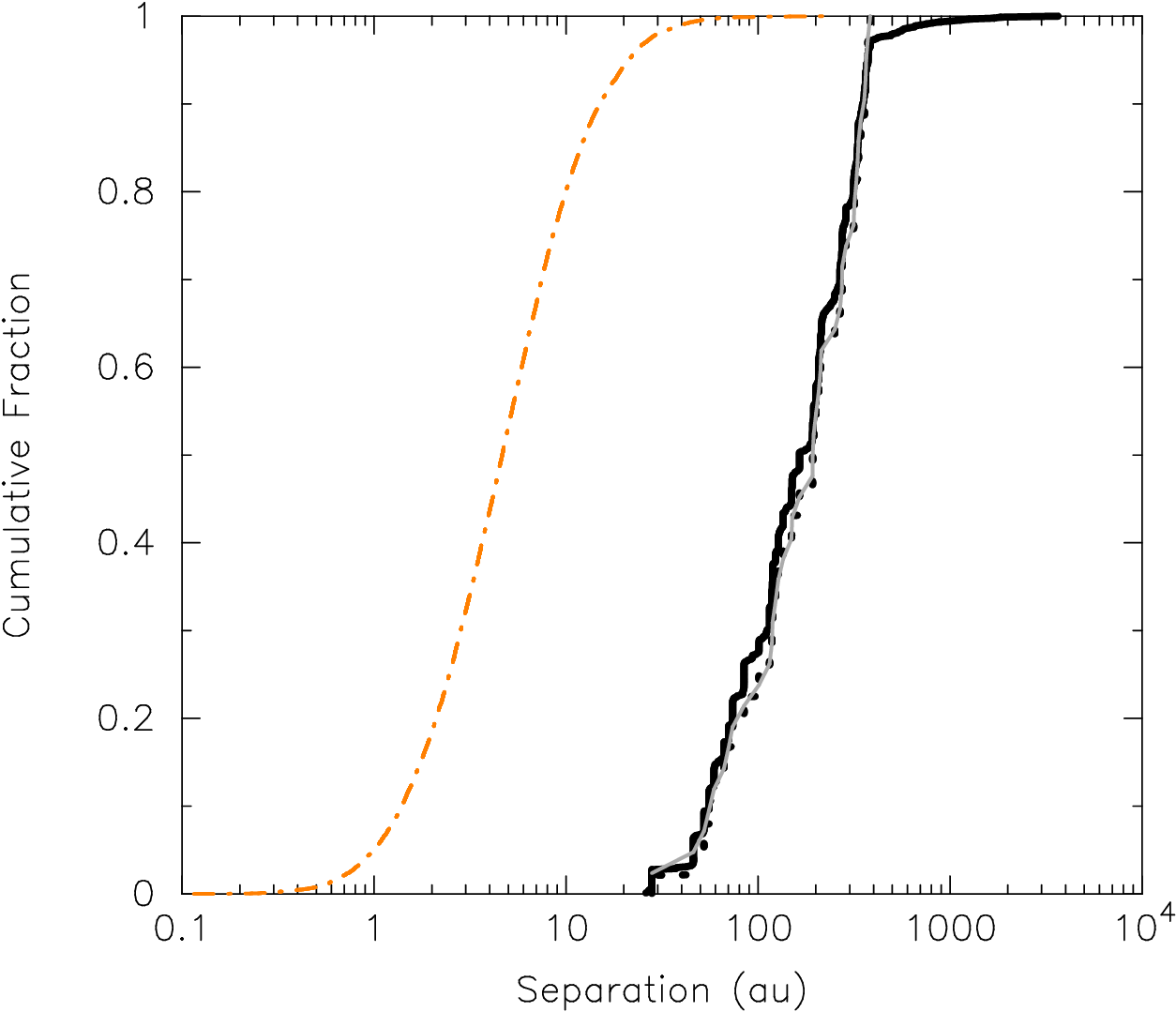}}}

\caption[bf]{The evolution of a planet--planet binary separation distribution where the separations are drawn from the observed JuMBOs separation distribution  for high (panel a) and low-density (panel b) simulations (simulations A and B, respectively). The black dotted line is the initial separation distribution, and the black solid line is the separation distribution after 1\,Myr, in the $N$-body simulations. The solid grey line is the observed JuMBO distribution \citep{Pearson23}, and the orange dot-dashed line is the fit to brown dwarf-brown dwarf binaries in the nearby Galaxy \citep{Basri06,Burgasser07,Thies07}. }
\label{fig:PP_sepcum}
  \end{center}
\end{figure*}

\begin{figure*}
  \begin{center}
\setlength{\subfigcapskip}{10pt}
\hspace*{-1.5cm}\subfigure[$\tilde{\rho} = 10^4$M$_\odot$\,pc$^{-3}$]{\label{fig:PX_sepcum-a}\rotatebox{270}{\includegraphics[scale=0.35]{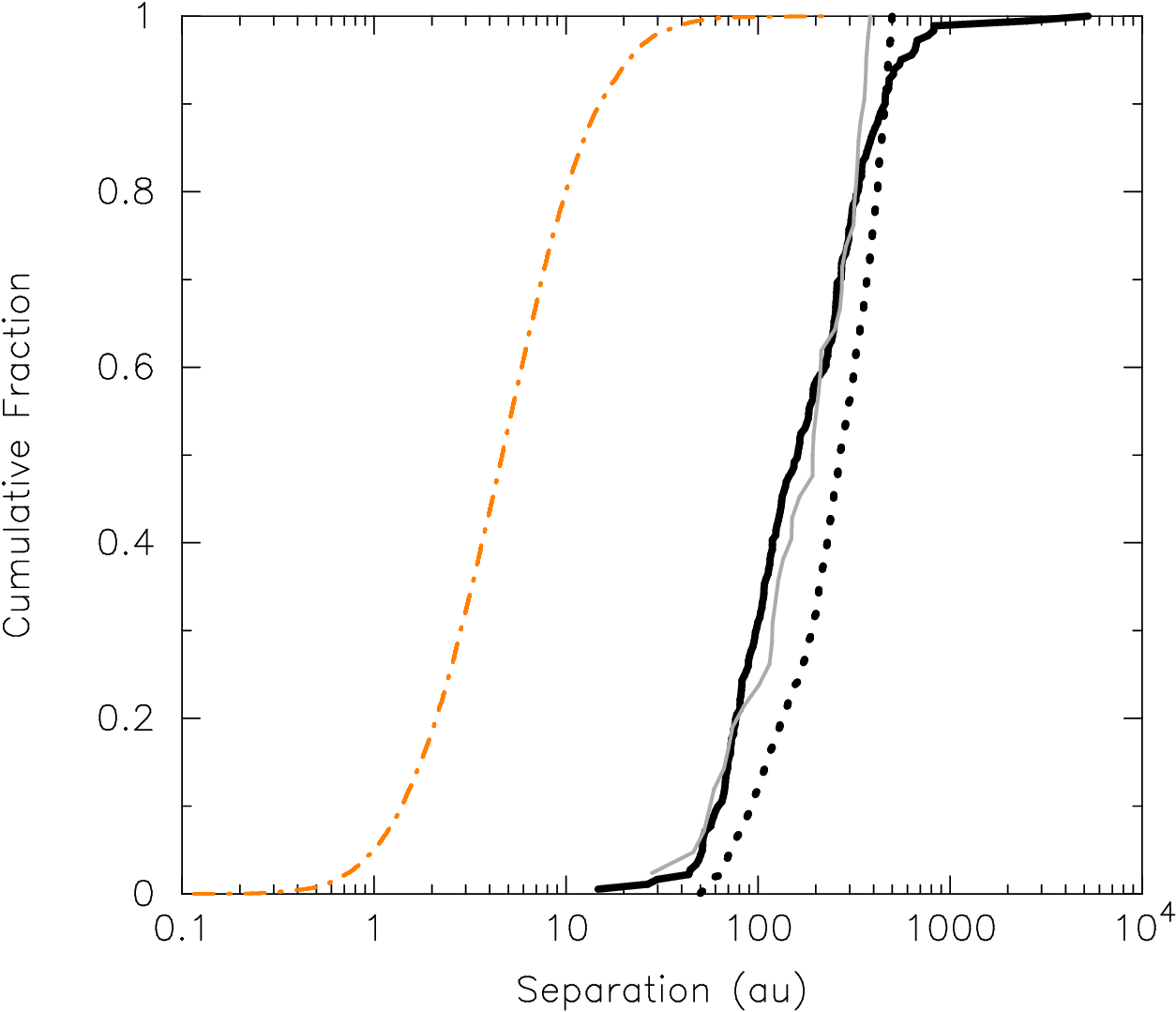}}} 
\hspace*{0.3cm} 
\subfigure[$\tilde{\rho} = 10^2$M$_\odot$\,pc$^{-3}$]{\label{fig:PX_sepcum-b}\rotatebox{270}{\includegraphics[scale=0.35]{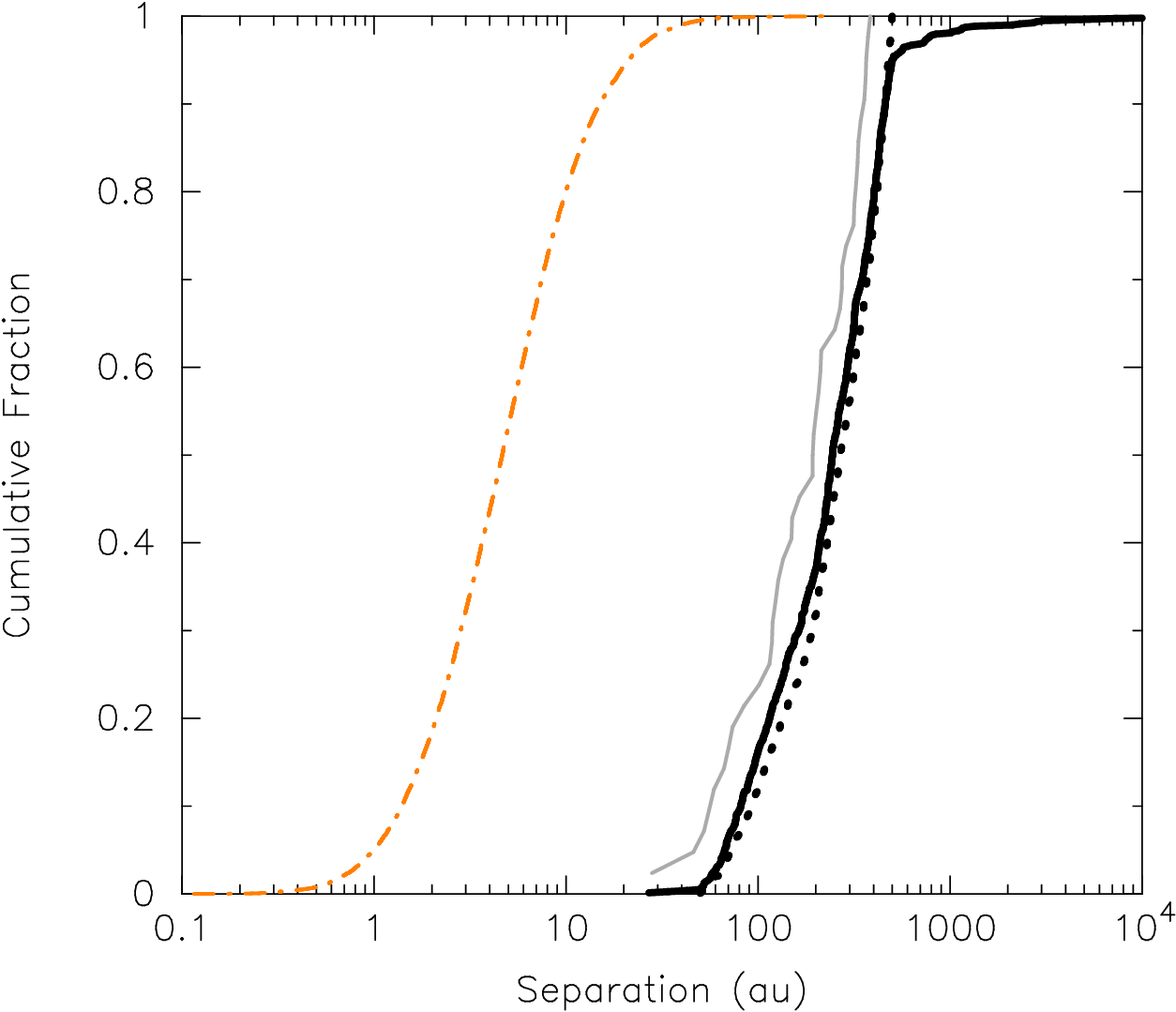}}}

\caption[bf]{The evolution of a planet--planet binary separation distribution where the separations are drawn from a flat distribution in the range 50 -- 500\,au for high (panel a) and low-density (panel b) simulations (simulations D and E, respectively). The black dotted line is the initial separation distribution, and the black solid line is the separation distribution after 1\,Myr, in the $N$-body simulations. The solid grey line is the observed JuMBO distribution \citep{Pearson23}, and the orange dot-dashed line is the fit to brown dwarf-brown dwarf binaries in the nearby Galaxy \citep{Basri06,Burgasser07,Thies07}. }
\label{fig:PX_sepcum}
  \end{center}
\end{figure*}

We can clearly see that the shape of the planet--planet separation distribution also changes due to dynamical interactions, with more wider systems being destroyed than the closer systems. To remove binning noise, we show the evolution of the separations as cumulative distributions in Figs.~\ref{fig:PP_sepcum}~and~\ref{fig:PX_sepcum}.

Fig.~\ref{fig:PP_sepcum} shows the evolution of the separation distribution for planet--planet binaries where the initial separations are taken from the observed distribution for the JuMBOs (the solid grey line). The initial distribution in the $N$-body simulations is shown by the dotted black line (which is statistically identical to the observed distribution), and the distribution after 1\,Myr is shown by the solid black line. For reference, the fit to the very low-mass and substellar binary separation distribution in the Galactic field is shown by the orange dot-dashed line \citep{Basri06,Burgasser07,Thies07}, which peaks at a mean separation of 4.6\,au, is normalised to a binary fraction of 0.15 and is valid for systems where the primary mass is $m_P \leq 0.1$\,M$_\odot$ and mass ratios $q > 0.1$. In panel (a) we show the results for initially dense ($\tilde{\rho} = 10^4$\,M$_\odot$\,pc$^{-3}$) star-forming regions (sim. A), and in panel (b) we show the results for lower density ($\tilde{\rho} = 10^2$\,M$_\odot$\,pc$^{-3}$) star-forming regions  (sim. B).

Clearly, if the stellar density during star formation in the ONC was similar to the present day (panel b), then the separation distribution does not change (but the overall fraction of planet--planet binaries will change, see the red dashed line in Fig.~\ref{fig:fbin}). However, if the ONC formed stars at higher densities than observed today, then the observed JuMBO distribution today cannot be the initial distribution, as many wider systems are destroyed, moving the overall distribution to shorter separations.

If the observed JuMBOs formed with a wider range of separations (e.g. 50 -- 500\,au), then in a dense star-forming region the destruction of more of the wider systems processes the initial population to shorter systems (Fig.~\ref{fig:PX_sepcum}, where the lines are as in Fig.~\ref{fig:PP_sepcum}). For this initial separation distribution, dynamical processing in dense regions  (sim. D, Fig.~\ref{fig:PX_sepcum-a}) would reproduce the observed JuMBO distribution, whereas dynamical processing in lower-density regions (sim. E, Fig.~\ref{fig:PX_sepcum-b}) would leave too many wider systems.

Regardless of the initial separation distribution, it is clear that a star-forming environment with a density similar to many nearby star-forming regions ($\geq 100$M$_\odot$\,pc$^{-3}$) would destroy many planet--planet binary systems with similar properties to the observed JuMBOs. This implies that even more systems than the 42 reported in \citet{Pearson23} would need to form, given the final binary fraction of 0.5 even in our lower-density simulations.

We also note that the observed distribution (the grey histogram in Fig.~\ref{fig:PP_sepdist}) shows an increasing trend to higher separations. This suggests that there may be even more wider JuMBOs that are not observed due to observational incompleteness, and these wider systems would be even more susceptible to dynamical destruction than the observed systems. If the data are incomplete, this implies that even more JuMBOs need to be produced by some formation mechanism(s) than our dynamical constraints suggest.

\section{Conclusions}
\label{conclusions}

We present $N$-body simulations of the evolution of star-forming regions in which we place a population of planet--planet binary systems with properties similar to the JuMBOs observed in the ONC \citep{Pearson23} to determine how many of these systems are affected by dynamical evolution in star-forming regions. Our conclusions are as follows:

(i) The relatively wide separations (10s to 100s au), combined with their low binding energies (due to their low masses) means that many planet--planet mass binaries are destroyed in our simulations. For the present-day density of the ONC, at least half of all systems are destroyed, but for the much more likely denser initial conditions for the ONC, up to 90\,per cent of these binaries are destroyed.

(ii) The implication of this high destruction rate is that to explain he observed population of 42 JuMBOs, a significantly higher number of primordial systems must have been present in the star-forming region, as many of these would be broken apart by dynamical encounters.

(iii) If the initial density of the ONC is high ($\sim 10^4$\,M$_\odot$\,pc$^{-3}$), then the observed JuMBO separation distribution has been dynamically sculpted, and the initial distribution would contain significant numbers of wider systems (up to 500\,au). If the inital density is lower ($\sim 100$\,M$_\odot$\,pc$^{-3}$) then the observed JuMBO separation distribution is similar to the initial separation distribution. \\

If the observed JuMBO population are planetary mass members of the ONC \citep[see][for an alternative explanation]{Luhman24a} then the primordial JuMBO systems must have been even more numerous than they are now, suggesting they are likely to be the end-point of photoerosion of the cores of objects that would otherwise have gone on to form the more commonly occuring stellar multiple systems \citep{Diamond24}. Such a scenario is supported by radio observations of candidate JuMBO~24 by \citet{Rodriguez25}, who find a poper motion velocity for this system of $<$6\,km\,s$^{-1}$, which \citet{Rodriguez25} argue is commensurate with the velocities of stars, rather than ejected planets \citep{Coleman24} \citep[although ejected planets can have proper motion velocities much lower than this,][]{Parker23c}.

\section*{Acknowledgements}

We thank the anonymous referee for a prompt and helpful report. RJP acknowledges support from the Royal Society in the form of a Dorothy Hodgkin Fellowship. 

\section*{Data availability statement}

The data used to produce the plots in this paper will be shared on reasonable request to the corresponding author.

\bibliographystyle{mnras}
\bibliography{general_ref}

\label{lastpage}

\end{document}